\documentclass[a4paper,11pt]{article}
\usepackage{pos}
\usepackage{wasysym}

\title{Recent developments of the SDHCAL prototype}

\author[1]{G\'erald Grenier}

\affiliation{IP2I,Université Lyon 1,IN2P3/CNRS\\
   Villeurbanne, France}

\emailAdd{gerald.grenier@univ-lyon1.fr}

\note{For the CALICE SDHCAL groups.}

\abstract{
After the construction and successful operation of 
the first technological prototype of the Semi-Digital Hadronic 
CALorimeter (SDHCAL), developed within the CALICE collaboration, 
new R{\&}D efforts have been initiated
to fully validate the SDHCAL option for future 
experiments proposed for the ILC and CEPC colliders.
The SDHCAL is a sampling hadronic calorimeter using large Glass Resistive Plate Chamber (GRPC) as 
active medium with embedded readout electronics. The GRPC prototype 
size is 1~m$^2$ while future detectors require GRPC detectors 
with scalable length up to 3~m long (0.9$\times$3~m$^2$). 
The readout Printed Circuit 
Board (PCB) consists of 1~cm$^2$ copper pads on one side and 
64-channel HARDROC readout chips on the other side.

The design of such large size scalable detectors has been addressed 
and has required rethinking the gas flow in the GRPC in order to 
maintain detection efficiency and spatial response homogeneity. The 
readout PCB was also redesigned to make it scalable in 
length and more tolerant of ASIC readout failures. 
It now uses the latest version of 
the HARDROC readout chips series. 
To deal with the maximum production size of a PCB with 8 layers, 
an ingenious scheme with several PCBs connected to each other by tiny, flexible connectors has been developped.
A new DAQ interface board with an optimized geometry to fit the 
requirements of the ILD detector, can handle a PCB area 
up to 2.76~m$^2$, sufficient to 
cope with the GRPC maximum size in ILD.
A new cassette, as part of the calorimeter absorber, is being 
designed. The main challenge is to ensure the rigidity and uniform 
contact between the GRPC and its PCB. For the ILC detector, the 
ILC beam time structure is used to power-pulse the ASICs to 
keep the power consumption low 
enough to avoid cooling the PCB. For the CEPC, the continuous 
operation of the accelerator implies adding cooling capacity to the 
designed cassette structures. In addition, tools to handle the new cassette are being finalized.
Finally, the way to manufacture the mechanical structure 
to support 3~m long GRPC with the needed improved flatness has 
been solved.
A first fully assembled prototype 
of 2~m$^2$ with 4 GRPCs is expected to be ready in year 2022.

In addition, new developpements to replace single gap GRPC by 
multigap GRPC coupled with fast timing electronics are being 
pursued. A time resolution better than 50 ps is achievable. This 
will allow to follow the temporal evolution of the hadronic showers 
developing in the calorimeter.
In parallel, the first SDHCAL prototype has been extensively tested 
in beam test facilities. Refined analysis techniques are being 
developed to improve the energy and shower reconstruction. The latest analysis developments cover techniques 
to improve the spatial uniformity of the response and a better 
treatment of the particle incidence angle in the energy 
reconstruction.
}

\FullConference{%
  *** Particles and Nuclei International Conference - PANIC2021 ***\\
  *** 5 - 10 September, 2021 ***\\
  *** Online ***
}

\begin{document}
\maketitle


\section{The SDHCAL}
The SDHCAL\cite{SDHCAL} is a stainless steel hadronic sampling calorimeter made with Glass Resistive Plate Chambers (GRPC) with embedded readout electronics as sensitive medium. 
It is the first technological prototype of a Particle Flow Algorithm (PFA) oriented calorimeter developped within the CALICE collaboration.
The SDHCAL has been designed for the ILD\cite{ILD} detector of the
International Linear Collider project (ILC)\cite{ILC}. For ILD, an ingenious mechanical structure avoiding cracks has been designed. 
This design implies variable length for the sensitive medium with a maximum size of 0.9$\times$3~m$^2$.
The challenges for this design are to obtain GRPC response homogeneity for large area, an active detector with a thickness of only a few mm including the embedded low power consumption electronics and all the services on one side of the detector.

\begin{figure}
\begin{minipage}[b]{0.49\textwidth}
\mbox{\includegraphics[width=\textwidth,trim=0 0 0 2cm,clip=true]{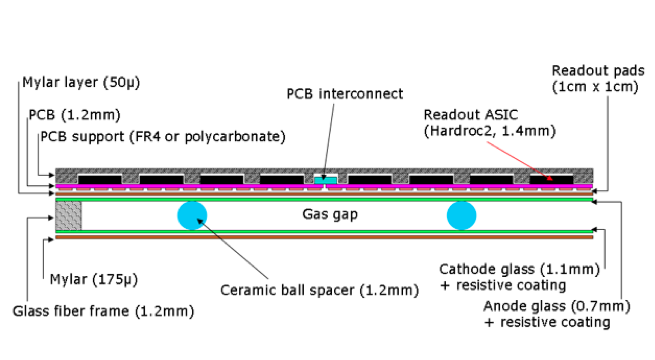}} \\
\includegraphics[width=\textwidth]{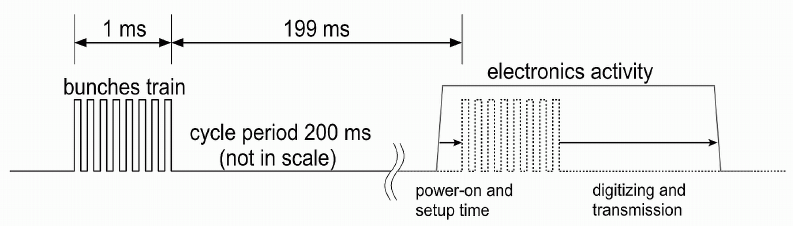}
\end{minipage}
 \includegraphics[width=0.49\textwidth]{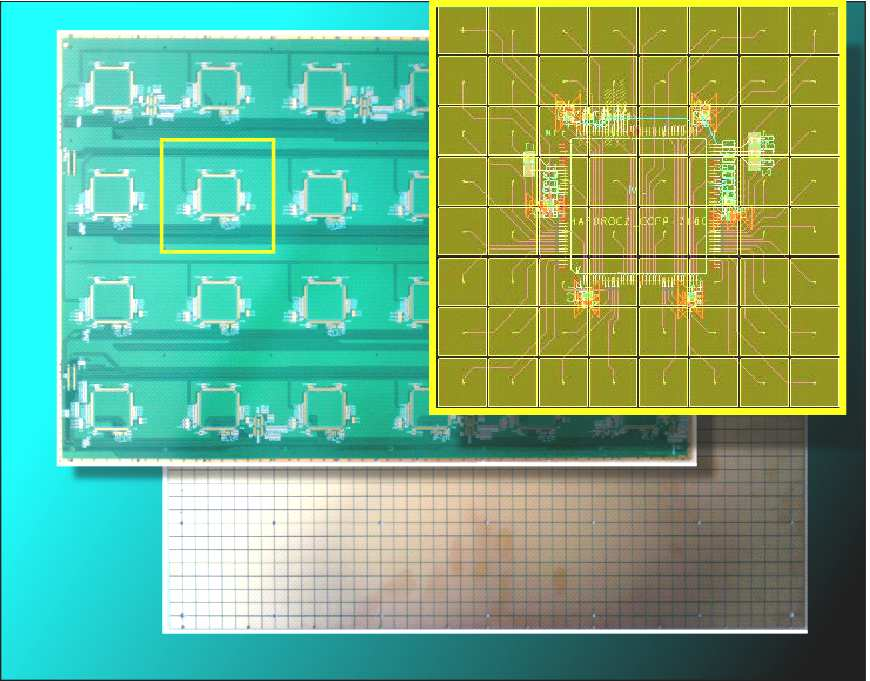}
 \caption{\label{fig:GRPC} Top left : Cross section view af a GRPC and its embedded electronics. 
 Bottom left : scketch of the ILC beam cycle.
 Right : View of the SDHCAL embedded PCB, one side with the HARDROC ASICs, the other side with copper pad, 
the inserted top right image shows the connections between the pads and the ASIC.}
\end{figure}

A technological prototype meeting all the above challenges has been conceived as a demon\-strator\cite{SDHCAL}. It is made with a stainless steel self-supporting structure able to hold up to 50 GRPC cassettes. 
A cassette is 11 mm thick and is a steel cover containing the GRPC and the embedded electronics. A cross section view of the 6~mm thick active element is shown on figure \ref{fig:GRPC}.
The electronics reads 1 m$^2$ GRPC using an array of 96$\times$96 copper pads of 1~cm$^2$ area. 
The 1~m$^2$ readout PCB is made by connecting six $\frac{1}{2} \times \frac{1}{3}$~m$^2$ PCB such as the one pictured on figure \ref{fig:GRPC}.
The full prototype has more than 460000 readout channels with less than $1\permil$ of them dead.
The readout electronics is made with daisy chained HARDROC2\citep{HARDROC} chips each reading 64 pads (8$\times$8 square) with a 3-level discriminator covering a dynamic range going from 10~fC to 30~pC
and producing a 2-bit (3 thresholds) readout per pad. 
To reach the low consumption target, the chip is used in power pulsing mode based on the ILC beam time cycle shown on figure \ref{fig:GRPC}. 
Each ASIC records signal during 1 ms storing up to 127 threshold crossing in internal memory. 
After 1 ms, it transmits the memory content and is switched off for nearly 200~ms and switched back on when new bunches train are coming.
Extensive tests have demonstrated that the power pulsing doesn't affect the detector performance when the ASIC is powered on at least 
25~$\mu$s before the first event to record.

\section{Test beam results}

\begin{figure}
\mbox{\includegraphics[width=0.345\textwidth,trim=9cm 0.8cm 2cm 2cm,clip=true]{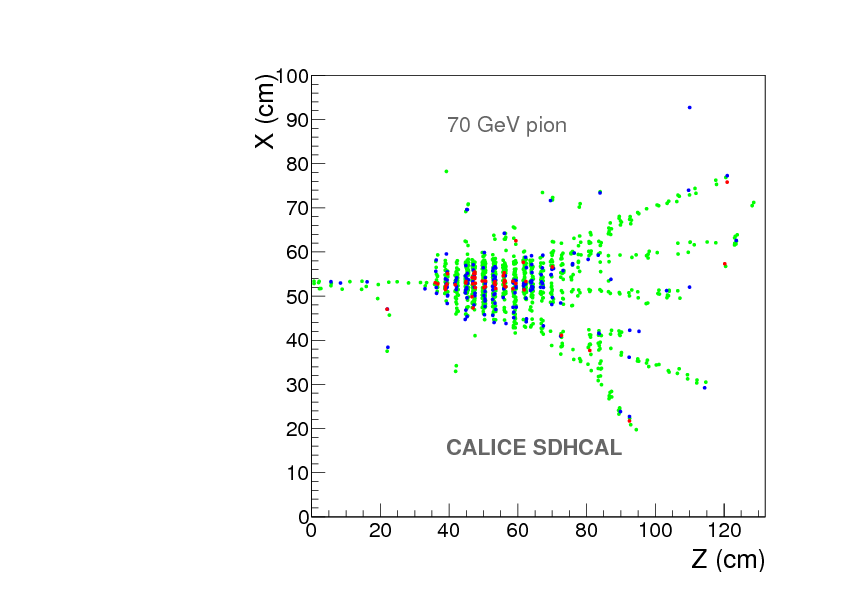}}
\includegraphics[width=0.39\textwidth]{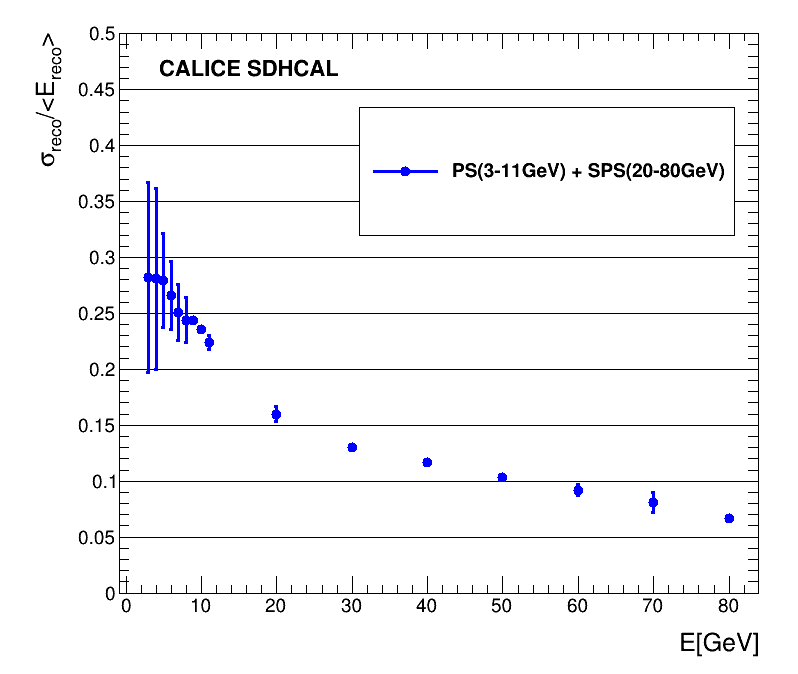}
\begin{minipage}[b]{0.25\textwidth}
\includegraphics[width=\textwidth]{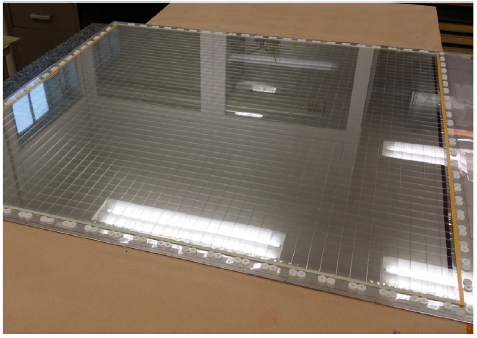}
\includegraphics[width=\textwidth]{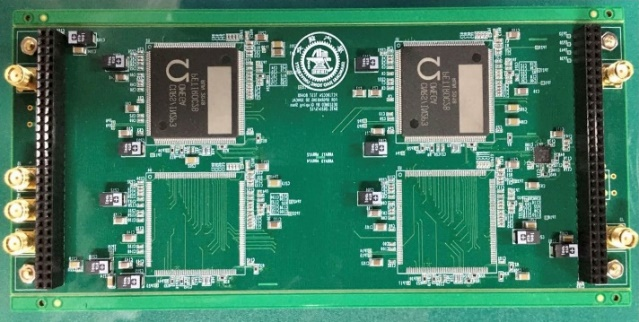}
\end{minipage}
\caption{\label{fig:testbeam} Left : event display of the interaction of a 70 GeV pion in the SDHCAL prototype. Color code correponds to the highest crossed threshold (first=green, second=blue, third=red). 
Middle : Resolution of the pion reconstructed energy as a function of the incoming pion energy. 
Left : prototype of a multigap GRPC (top) and of a PCB using PETIROC chip (bottom).}
\end{figure}

The SDHCAL prototype has been tested under various hadron beams at the CERN accelerators \cite{TestBeam}. The figure \ref{fig:testbeam} left shows a pion shower recorded in the SDHCAL prototype.
Each point is a pad crossed by particles and the color code corresponds to the highest threshold crossed. 
The threshold values were set to 110~fC, 5~pC and 15~pC corresponding qualitatively to one, few and many charged particles crossing the pad. 
The incoming particle energy is estimated as 
\begin{equation}
E_{reco} = \alpha N_1 + \beta N_2 + \gamma N_3
\label{eq:Ereco}
\end{equation}
where $N_i$ is the number of fired pads with $i$ as the highest crossed threshold 
and $\alpha$, $\beta$, $\gamma$ are quadratic polynomial functions of $N_{tot}$, the total number of fired pad, $N_{tot} = N_1 +N_2+N_3$. The relative deviation of the estimated energy is less than 5 \% of the true energy and the resolution in energy is shown in figure \ref{fig:testbeam} middle.
Recently by combining advanced reconstruction techniques like counting the tracks in shower using Hough transform\cite{Hough}
with Boosted Decision Tree, selection of high purity pion shower 
has been achieved down to 3~GeV, allowing to test the prototype performance in energy resolution on the low energy side\cite{BDT} 
as is shown on figure \ref{fig:testbeam} middle.

The SDHCAL prototype has also been exposed to pion beam with an incident angle $\theta$ different from zero. 
The number of fired pads are increased by a factor $\frac{1}{\cos \theta}$ due to geometrical effects. 
Preliminary study tends to show that replacing the $N_i$ by $N_i \cos \theta$ in equation \ref{eq:Ereco} is sufficient to keep the energy reconstruction performance at the same level as the case of normal pion incidence.  
By adjusting the HARDROC ASIC individual thresholds and gains, 
the response uniformity to crossing muon can be significantly 
improved both within 1 GRPC and between different GRPCs. 
Effects of such tuning on the shower energy reconctruction is 
currently being studied. 

\section{New R{\&}D}

The current R$\&$D activities on the SDHCAL follow two paths. 
Finalising the proposal for the ILD detector and adapting the prototype to other future leptonic colliders.

The ILD design requires to built variable lengths GRPC with a maximum length up to 3~m.
To achieve this, prototypes of 2~m long GRPCs have been built 
with a gas distribution system inside the chamber that is length scalable. 
A mechanical structure able to host GRPC of size up to 1$\times$3~m$^2$
has been build. To keep the aplanarity of the GRPC cassette holding slots below 1~mm, the absorber plates had to be processed by 
roller leveling and assembled with a sophisticated method of joining the 
plates using electron beam welding techniques since usual screw and bolt assembly is not sufficient. 
In parallel, all the electronics has been redesigned. The PCBs hosting the 1~cm$^2$ pads have a size of 1$\times\frac{1}{3}$~m$^2$
and can be chained at their larger side. 
They are equipped with the new HARDROC3 chip which has an extended dynamic range up to 50~fC. 
The PCB clock 
distribution, slow control and fast control have been redesigned to 
speed up communication over long chains of ASICs and PCBs. 
The clock is distributed via the TTC protocol while the slow and 
fast controls use I$^2$C protocol.
A new Detector Interface Board has been designed and is able to communicate directly through I$^2$C links with up to 432 ASICs. 
These characteristics are enough to handle the maximal GRPC size foreseen for ILD.

In parallel, water cooling system for the SDHCAL is designed 
to make the calorimeter usable in circular collider like CEPC\cite{CEPC} where the power pulsing can't be used.
The last R$\&$D program is to replace the GRPC by multigap GRPC (MGRPC) and the HARDROC chip by PETIROC\cite{PETIROC}. The current prototypes 
are displayed on figure \ref{fig:testbeam} right. The MGRPC 
efficiency has been measured to be 94\%\cite{MGRPC}.
The PETIROC chip can measure both energy and time with a resolution better than 20~ps. Combined with the time resolution of multigap GRPC, the time resolution for each crossing pads could be below 50~ps. 
Such equipped, the SDHCAL prototype can become a space-time imaging hadronic calorimeter fitted for 5D calorimetry.


\end{document}